\documentclass[runningheads]{llncs}
\usepackage{graphicx}
\usepackage{dirtytalk}
\usepackage{paralist}
\usepackage{tcolorbox}
\usepackage{float}
\usepackage{cite}
\usepackage{orcidlink}
\usepackage{tabularx}
\usepackage{caption}
\usepackage{multirow}
\usepackage{algorithmic}
\usepackage{rotating}
\usepackage{amsmath,amssymb,amsfonts} 
\usepackage{booktabs}

\usepackage{array}
\usepackage{arydshln}
\let\llncssubparagraph\subparagraph
\let\subparagraph\paragraph
\let\subparagraph\llncssubparagraph
\begin{document}

\title{
User-Like Bots for Cognitive Automation: \\
    A Survey
    }
\author{
Habtom Kahsay Gidey\inst{1}\orcidlink{0000-0001-5802-2606} \and
Peter Hillmann\inst{1}\orcidlink{0000-0003-4346-4510} \and
Andreas Karcher\inst{1} \and
Alois Knoll\inst{2}\orcidlink{0000-0003-4840-076X}
}
\authorrunning{H.K Gidey et al.}
%
\institute{
Universität der Bundeswehr München, Germany \\
\email{\{habtom.gidey, peter.hillmann, andreas.karcher\}@unibw.de} \and
Technische Universität München, München, Germany\\
\email{\{knoll\}@in.tum.de}
}
\maketitle              
\begin{sloppypar}
\begin{abstract}
Software bots have attracted increasing interest and popularity in both research and society. 
Their contributions span automation, digital twins, game characters with conscious-like behavior, and social media. 
However, there is still a lack of intelligent bots that can adapt to the variability and dynamic nature of digital web environments. 
Unlike human users, they have difficulty understanding and exploiting the affordances across multiple virtual environments.

Despite the hype, bots with human user-like cognition do not currently exist.
Chatbots, for instance, lack situational awareness on the digital platforms where they operate, preventing them from enacting meaningful and autonomous intelligent behavior similar to human users.

In this survey, we aim to explore the role of cognitive architectures in supporting efforts towards engineering software bots with advanced general intelligence. 
We discuss how cognitive architectures can contribute to creating intelligent software bots.
Furthermore, we highlight key architectural recommendations for the future development of autonomous, user-like cognitive bots.
\end{abstract}

\section{Introduction}\label{intro}

Software bots are becoming an integral part of automation and social computing. 
Digital platforms, including software ecosystems and cyber-physical systems, are growing increasingly complex. 
The complexity of diverse digital systems can overwhelm even expert human users~\cite{jiang2018cognitive}.                
In such scenarios, software agents acting as autonomous users can assist in automating human user activities. 
As a result, there is a growing interest in augmenting bots into software-intensive business, social, or industrial environments for cognitive automation~\cite{engel2022cognitive}.
In Industry 4.0 (I4.0) and digital twins (DTs), software agents are playing a crucial role in enabling smart factories with higher flexibility, efficiency, and safety~\cite{leitao2016smart,lee2018industrial,karnouskos2020industrial}. 
Additionally, in the service industry, robotic process automation (RPA) leverages bots to automate business processes~\cite{ivanvcic2019robotic}.

Software development bots, also known as DevBots, are making their mark in automated software engineering~\cite{erlenhov2019current,wessel2018power}. 
It has become common to see bots assisting in code review and bug-fixing on platforms like GitHub~\cite{monperrus2019explainable,wessel2018power}. 
Wessel et al.~\cite{wessel2018power} identified 48 bots used for this purpose. 
From committing code to coordinating open-source projects, bots are increasingly becoming a part of the software development life cycle~\cite{monperrus2019explainable,wessel2018power}. 
Their impact on development can significantly affect how future digital innovation ecosystems are managed and governed~\cite{newton2022everyboty,platis2021software,maruping2020governance}.
Social platforms and games also serve as environments for social bots and virtual avatars~\cite{hendler2010semantic,arrabales2009towards}.
Although claims of political intent and the influence of social bots on social media are exaggerated, bots are also prevalent on social platforms nowadays~\cite{gallwitz2021rise}.
The diverse applications of bots highlight the desiderata and requirements for advanced cognitive agents~\cite{mcdonald2019cognitive}.

However, the reality falls short of the expectations, and advanced social bots do not exist today~\cite{gallwitz2021rise,butlin2023consciousness}. 
First, the level of autonomy in industrial software agents and robotic process automation (RPA) is minimal~\cite{kugele2021towards,vagia2016literature}. 
Agents often have architectures that are tightly coupled with specific service platforms.
As a result, bots are designed and optimized for these specific platforms, limiting their adaptability. 
They lack the ability to autonomously recognize the variability and then function effectively across diverse web or service environments.
Consequently, they lack a sense of awareness and the ability to adapt to different contexts.
Second, bots have very limited or no autonomy in their behavior. 
Addressing these challenges requires a focus on identifying and resolving architectural concerns~\cite{baldassarre2017architecture,kraetzschmar2018software,bosch2021engineering,goertzel2014engineeringPart1}. 
It is important to note that bots in social media, games, industry, and business use cases may differ, but the software systems architecture and engineering challenges can span domains~\cite{wooldridge1998pitfalls,martinez2022software}.

Thus, by reviewing and synthesizing existing works, this study aims to explore two architectural aspects of bots through two research questions.
The first question investigates the role of cognitive architectures in bot behavior, while the second question examines the strict separation of bots and their operational environment. 
In our context, the second question is operationalized by evaluating \emph{user-likeness} or similarity. 
A \emph{user-like bot} refers to the level of similarity a bot possesses compared to a human user~\cite{gidey2023towards}.

The subsequent sections are structured as follows: Section~\ref{background} provides the foundational background on bots and cognition, Section~\ref{method} describes the approach employed in surveying bots, Section~\ref{results} presents the evaluated results, Section~\ref{discussion} discusses the implications of the obtained results, and finally, Section~\ref{conclusion} concludes the study.

\section{Bots and Cognition}\label{background}

\subsection{Software Bots}\label{bots}

Software bot, or simply bot, is an umbrella term for diverse software agents~\cite{lebeuf2017software}.
The term is used loosely in domains such as RPA. 
RPA, robotic process automation, is an automation approach that employs software bots, sometimes referred to as digital workers~\cite{ivanvcic2019robotic}.
RPAs and digital twins utilize various toolsets and development paradigms from agent-based systems~\cite{karnouskos2020industrial,wooldridge1995agent}.

Lebeuf~\cite{lebeuf2018taxonomy} conducted a comprehensive study proposing a broad definition and general taxonomy of bots. 
She defines software bots as interfaces that connect users to software services and describes them as a~\say{\textit{new [user] interface paradigm}}.
According to this perspective, users can access software services through a bot, where the user interface takes the form of a conversational interface.
The user is typically human, although other programs and systems can also utilize the bot.
Software services, in this context, refer to applications or digital platforms that provide additional functionalities.
While these services are typically external, they can also be integrated as internal components of the bot.

Furthermore, Lebeuf~\cite{lebeuf2018taxonomy} classifies bots based on their observable properties and behaviors. 
The taxonomy defines three dimensions: environment, intrinsic, and interactions.
The environment dimension refers to the software service properties that bots operate on.
Ideally, a software bot is separate from a specific platform and can operate on multiple platforms.
However, most bots are tailored to a specific platform, such as a Twitter bot. 
The intrinsic dimension describes the internal abilities of the bot. 
Lebeuf also puts anthropomorphism as an intrinsic dimension, which specifies if the bot has human~\emph{user-like} features, such as name, visualization, and persona. 
The interaction dimension specifies how the bot accesses and interacts with the environment.

Lebeuf's taxonomy is based solely on observable properties and does not consider the architectural aspects and components of bots~\cite{lebeuf2018taxonomy}.
In our context, the interaction dimension concerns perception, action, and autonomy, which impact the system architecture and environment of the bot.
Therefore, in this study, we introduce another term to restrict the scope to bots that are similar to users.

\subsubsection{User-like Software Bots}

Today, the web is an operational environment for human users and software user agents~\cite{gidey2023towards}. 
Unlike virtual reality environments, where agents are typically manifested as conscious-like avatars, the web represents a different kind of mixed-reality environment. In this environment, both human users and software bots interact, tying virtual elements with real-world extensions, as exemplified by the Web of Things (WoT)~\cite{holz2011mira,milgram1994taxonomy,charpenay2022unifying}.
A diverse set of software agents interact with the web as their environment~\cite{charpenay2022unifying,lebeuf2018taxonomy}.

In this study, we refer to these classes of mixed-reality software agents as~\emph{user-like bots}~\cite{gidey2023towards}. 
The term~\emph{user-like} implies that these bots exhibit similarities to human users. 
User-like bots use graphical user interfaces to perceive and act within a software service environment.
They interact with keyboard and mouse operations.

As Gibson's theory of visual perception suggests, perception is not merely about passive observation; it's also about actively distinguishing the potential actions or~\say{\textit{affordances}} that an environment offers to an agent~\cite{gibson1977theory}.
The environment, in this context, presents affordances~\cite{nye2012affordances}.
Affordances refer to possibilities for interaction and action~\cite{gidey2023towards}.
Platform services or features can be analogous to affordances in the real world.
User-like bots are expected to understand or perceive these affordances and act on them similar to human users. 

While the web is the primary context, a software service environment can also refer to any desktop application.

\subsection{Cognitive Architectures}
Research in engineering machine intelligence, particularly Artificial General Intelligence (AGI), aims to endow software systems with cognitive abilities that enable machines to think at or beyond the level of humans~\cite{goertzel2014engineeringPart1}.
To achieve this goal, efforts to understand the brain from disciplines such as cognitive and neuroscience have led to various promising approaches.
One such approach is the study of cognitive architectures, which focuses on designing high-level cognitive functions~\cite{gidey2023towards,vernonCAs2022}.

Cognitive architectures serve as essential architectural design foundations for artificial intelligence research. 
The ultimate objective of cognitive architectures is to enable software with cognitive abilities equal to or greater than human-like intelligence.
Cognitive architectures can be described as a set of specifications or theories of cognition that outline the essential structural elements and capabilities of a cognitive system~\cite{vernon2014artificial,kotseruba2016review}.
Metzler and Shea~\cite{metzler2011taxonomy} compiled a list of cognitive functions as components that constitute a cognitive architecture, such as learning, reasoning, decision-making, perception, planning, and acting.

Cognitive architectures are designed to handle a broader set of cognitive tasks or cognitive functions. 
They can enable perpetual learning from the environment, adapting to changes, and reasoning based on available information. 
Due to this universal approach, an agent implementing a cognitive architecture may operate successively and simultaneously in various applications. 
While most cognitive architectures remain theoretical specifications, some, such as ACT-R, LIDA, and SOAR, have implementations and active communities~\cite{vernon2014artificial}.
Duch et al.~\cite{duch2008cognitive} have conducted an in-depth comparison on a technical level of these architectures, among others.
\begin{figure}[htbp]
\centerline{\includegraphics[scale=0.2]{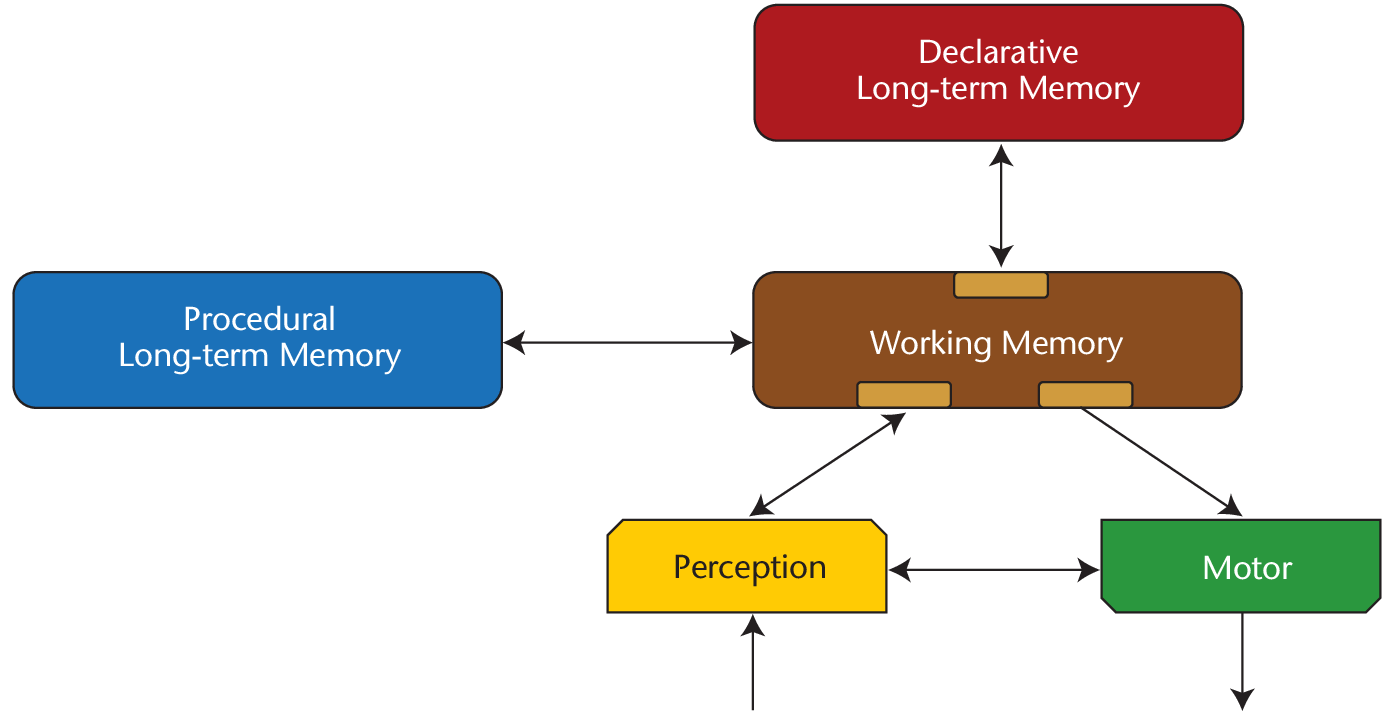}}
\caption{Schematic Structure of the Standard Model of Cognition~\cite{laird2017standard}.}
\vspace*{-0.5cm}
\label{fig:genericModelCA}
\end{figure}

Currently, efforts are being made in the cognitive modeling research community to establish a comprehensive understanding of the architectural assumptions that define aspects of human-like cognition, whether natural or artificial~\cite{laird2017standard}.
Fig.~\ref{fig:genericModelCA} shows a schematic structure of the Standard Model of Cognition~\cite{laird2017standard}.
The Standard Model of Cognition aims to consolidate knowledge from all existing cognitive architectures and establish a unified understanding of the generic aspects of cognition, such as perception and motor functions, common to all cognitive architectures. 

\subsection{Cognitive Automation}\label{models}

Traditional automation methods and approaches, such as process automation and RPA, have played crucial roles in automating repetitive tasks and workflows~\cite{van2004workflow,van2018robotic,engel2022cognitive}.
However, as organizations strive for increased agility, efficiency, and even hyper-automation, the increased complexity of cyber-physical systems necessitates additional layers of automation, i.e., cognitive automation~\cite{engel2022cognitive}.

Cognitive automation aims to advance the capabilities of traditional automation and RPA by combining them with other technologies in artificial intelligence. 
This integrated approach enables the automation of more complex and cognitive tasks that traditionally require human intervention.
Cognitive automation aims to automate knowledge and service work that involves decision-making, problem-solving, and other cognitive activities~\cite{engel2022cognitive,bruckner2011cognitive}.
It focuses on alleviating the burden of cognitive tasks on humans by automating their roles or enhancing mixed reality collaborations~\cite{van2018robotic,aguirre2017automation,engel2022cognitive}.

Similarly, advanced user-like bots or software agents can be employed to achieve generalizable intelligence or cognitive capabilities on digital platforms comparable to human users~\cite{gidey2023towards}.
As autonomous users, software agents can then assist in automating human user activities.
Consequently, one prominent application of cognitive automation can be utilizing user-like bots. 
Bots with generalizable intelligence, matching human users, are one way to address the augmentation of autonomous digital workers to cyber-physical systems. 
These bots can enhance efficiency in various domains, such as RPA, knowledge platforms, digital twins, and smart factories~\cite{karnouskos2020industrial,gidey2023towards}.

\section{Research Approach}\label{method}

In this section, we describe the study method and its execution. 

\subsection{Planning}\label{planning}

We conducted a literature review to examine the integration of cognitive architectures into the engineering of user-like software bots.
This approach allows us to investigate and challenge existing architectural assumptions of software bot development, which fail to enact meaningful, intelligent behavior. 
The method also uses specific criteria to narrow the review's scope and establish a clear and replicable methodology.
The survey is executed by systematically searching, selecting, and evaluating relevant works. 

\subsection{Initial Selection}\label{Selection}

Initially, we collected various bots, agents, software tools, and personal digital assistants to get an overview. 
Since terms such as agent, bot, software bot, user-agent, and chatbot are interchangeably used in literature, the initial efforts resulted in an extensive collection of sources. 
However, this initial process helped explore and map the software bots already developed throughout all time.

We then limited the collection to the bots that use some form of cognitive architecture or cognitive model. 
Furthermore, we determined the search to include works that distinctively show some similarity to a way a human user would access and with software service platforms, hence only user-like software bots. 

First, we systematically selected relevant works from comprehensive databases, such as Scopus.
To that end, specific keywords, including variations of~\say{software bot,}~\say{cognitive architecture,} and~\say{user-like or user-agent,} were devised and conducted.
Next, we used forward and backward snowball sampling~\cite{wohlin2014guidelines} to find citation chains of relevant studies that claim the development of bots with cognitive capabilities. 

Our initial collection resulted in approximately 190 works. With a closer look, we found that many were unrelated to the distinct interest. 
The initial results are then narrowed to a representative selection of software bots for further investigation and evaluation.

\subsection{Selection and Evaluation Criteria}\label{EvaluationCriteria}

\begin{figure}[htbp]
\centerline{\includegraphics[scale=0.16]{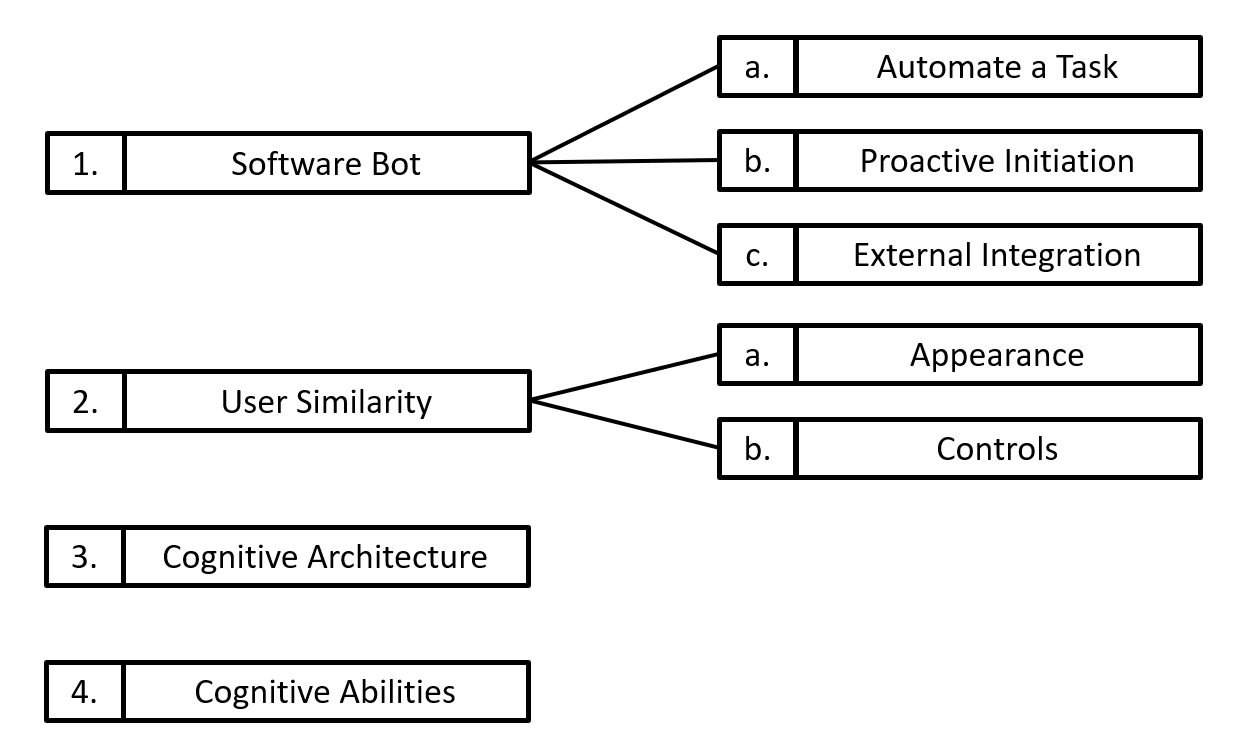}}
\caption{Criteria used to evaluate the bots applying cognitive models.}
\vspace*{-0.5cm}
\label{fig:criteria}
\end{figure}

The representative selection, also called candidates, is evaluated with four essential selection criteria. 
The four criteria used to evaluate the selection are derived to ensure the proper implementation of the interest and objective of the study. 
Each candidate bot selected is evaluated against all four criteria and the corresponding sub-criteria. 
Fig.~\ref{fig:criteria} depicts the four criteria and subcriteria of the first two: Software Bot and User Similarity.

\textit{(1) Software bot:} this criterion evaluates bot capability. It helps exclude conversational interfaces or chatbots that do little or no autonomous action on service platforms. 
The criterion has three subcriteria: a. Automate a Task, b. Proactive Initiation, and c. External integration.
One of the three aspects is the necessity that the bots automate a task in a digital environment (a. Automate a Task). 
In this context, the scope of the environment and the complexity of the task are neglected. 
To differentiate the candidate from other programs and scripts, as a second aspect, independence from the user is required (b. Proactive Initiation). 
Changes to other elements of the environment, combined with intrinsic motivations, should trigger the behavior and actions the bot performs. 
Indirect reactions to the actions of the user are permitted. 
For the third aspect, the candidate bot should operate externally in the targeted digital environment (c. External integration).
Direct integration into the environment, as a fixed part of a system, is not considered user-like. 

\textit{(2) User Similarity:} Equal to the first criterion, this criterion stems from the distinction set regarding user-like software bots in Sect.~\ref{background}. 
User-like is used to denote agents who either resemble human users or act on behalf of users. 
We focus on drawing a clear distinction between the bots that autonomously enact user-like behavior and others. 

\textit{(3) Cognitive Architecture:} While claiming the implementation of a cognitive architecture is relatively easy, its practical application poses a challenge. 

\textit{(4) Cognitive Abilities:} Implementing a cognitive architecture does not necessarily result in cognitive behavior. 
Consequently, this criterion unites various aspects regarding a candidate's cognitive abilities. 
The list aims at excluding narrowly set low-level heuristics from the selection while considering the desired high-level cognitive functions or generalized cognitive capabilities. 

\section{Results}\label{results}

This section presents the selected agents, personal assistants, and applied cognitive models. 
Their evaluation is based on the criteria outlined in Sect.\ref{method}. 
Furthermore, descriptions of notable selected results from the corresponding review are provided. 
Table~\ref{evaluation} shows the main results of the software bots analyzed and evaluated. The entries are ordered by the year of the publication. 
However, other arrangements are also feasible. 
The meaning of the encoded column captions is provided in Sect.~\ref{EvaluationCriteria} and Fig.~\ref{fig:criteria}. 
The elements of the selection cover diverse research domains. 
In the following, we highlight the significant bases of the elements in Table~\ref{evaluation}. 
It is worth noting that none of the candidates in the related publications are described as~\say{\textit{user-like software bots}} as termed in this study.
They were included because their implicit descriptions in the literature align with the previously set criteria and guidelines.

\begin{sidewaystable*}[htbp]
\centering
\caption{Overview of selected bots, agents, or assistants, with evaluation results.}
\label{evaluation}
\resizebox{\textwidth}{!}{%
\begin{tabular}{p{0.10\textwidth}p{0.36\textwidth}p{0.02\textwidth} p{0.02\textwidth}p{0.02\textwidth}p{0.04\textwidth}p{0.04\textwidth}p{0.32\textwidth}c} \hline
{\textbf{Study Ref.}} &
  {\textbf{Description}} &
  \multicolumn{3}{l}{\textbf{\begin{tabular}[c]{@{}l@{}}Software \\ Bot\end{tabular}}} &
  \multicolumn{2}{l}{\textbf{\begin{tabular}[c]{@{}l@{}}User \\Similarity\end{tabular}}} &
  {\textbf{Cognitive Architecture}} &
  {\textbf{\begin{tabular}[c]{@{}l@{}}Cognitive \\ Abilities\end{tabular}}} \\ \cline{3-7} &
   &
  \textit{\textbf{a.}} &
  \textit{\textbf{b.}} &
  \textit{\textbf{c.}} &
  \textit{\textbf{a.}} &
  \textit{\textbf{b.}} &
   &
   \\ \hline
Franklin et al. 1996~\cite{franklin1996virtual} & Virtual Mattie is a clerical agent that collects and shares weekly email information with a designated group. &  \checkmark & \checkmark & \checkmark & \checkmark & \checkmark & \checkmark . Extended versions of Maes' behavior net~\cite{maes1990situated} and Hofstadter and Mitchell's Copycat architecture~\cite{hofstadter1994copycat}. & X \\ \hdashline 
Franklin et al. 1998~\cite{franklin1998ida} & IDA: Intelligent distribution agent for US Navy sailor billet assignments via email. & \checkmark & \checkmark & \checkmark & \checkmark & \checkmark & \checkmark. Architecture based on global workspace theory. The basis for LIDA~\cite{franklin2012global}. & \checkmark \\ \hdashline 
Knoblock et al. 2003~\cite{knoblock2003deploying} & A personal information agent that gathers information from sources on the internet and links them to the specific task. & \checkmark & \checkmark & \checkmark & X & X & \checkmark. Employs, in combination with others, the rule-based approach, Planing by Rewriting~\cite{ambite2001planning}. & \checkmark \\ \hdashline 
Berry et al. 2004~\cite{berry2004personalized} & PCalM is a personalized calendar agent that manages the calendar of an individual and coordinates as well as schedules meetings. & \checkmark & \checkmark & \checkmark & X & X & \checkmark. Utilizes the Open Agent Architecture to connect the cognitive components directly~\cite{cheyer2001open}. & \checkmark \\ \hdashline 
Freed et al. 2008~\cite{freed2008radar} & RADAR is a personal assistant that reduces email overload by identifying task-relevant content and managing emails. & \checkmark & \checkmark & \checkmark & X & X & X. Interconnected structure of task-specific, AI-enhanced modules/agents. & X \\ \hdashline 
Berry et al. 2011~\cite{berry2011ptime} & PTIME is a learning cognitive assistant agent that represents a personalized time manager. & \checkmark & \checkmark & \checkmark & X & X & \checkmark. Employs a preference module, which is connected to various schedulers. & \checkmark \\ \hdashline 
SRI International~\cite{apple2021siri} & CALO is an extensive cognitive agent that learns and organizes. It is a predecessor of Apple Siri. & \checkmark & \checkmark & \checkmark & X & X & \checkmark. Cooperating cognitive agents including, for example, PTIME for language processing, see~\cite{tur2010calo}. & \checkmark \\ \hdashline 
Strain et al. 2014~\cite{strain2014learning} & Medical Agent X, an agent for clinical diagnostics. & \checkmark & X & X & X & X & \checkmark. Employs the cognitive architecture LIDA~\cite{franklin2012global}. & \checkmark \\ \hdashline 
Lebiere et al. 2015~\cite{lebiere2015functional} & An automated approach to determine the task of a piece of malware. & \checkmark & X & X & X & X & \checkmark. Partially employs the cognitive architecture ACT-R~\cite{ritter2019act}. & X \\ \hdashline 
Shi et al. 2017~\cite{shi2017world} & An experimental platform where agents learn to interact with web environments using only keyboard and mouse operations. & \checkmark & \checkmark & \checkmark & X & \checkmark & X Training models based on supervised and reinforcement learning~\cite{shi2017world}. & X \\ \hdashline 
Wendt et al. 2018~\cite{wendt2018usage} & A cognitive agent for building energy management. & \checkmark & \checkmark & \checkmark & X & X & \checkmark. Employs the KORE cognitive architecture for building automation~\cite{zucker2016cognitive}. & \checkmark \\ \hline
\end{tabular}
}
\end{sidewaystable*}

\subsection{IDA and Virtual Mattie}

The agent Virtual Mattie was developed by Franklin et al. in 1996 to perform clerical tasks related to the organization of seminar schedules~\cite{franklin1996virtual}. 
Human seminar organizers are contacted via email. 
The primary cognitive abilities of Virtual Mattie are aimed at understanding the free-form and probably incomplete messages. 
While Virtual Mattie implements mechanisms for goals and attention, the small number of cognitive abilities does not meet the respective criteria. 
The other criteria are satisfied.

IDA, the intelligent distribution agent designed by Franklin et al. in 1998 for the US Navy, is the successor of Virtual Mattie and the predecessor of the cognitive architecture LIDA~\cite{franklin2006lida}. 
The agent assigns new long-term tasks to sailors who have finished their current ones. 
A notable feature of the architecture is its working memory, based on the global workspace theory~\cite{baars2005global}. 
In comparison to Virtual Mattie, IDA possesses more enhanced cognitive abilities. 
In addition to decision-making and attention modules, IDA also includes a module for emotions. 
Due to the communication by email and similar related access to human coworkers, IDA, at a basic level, satisfies the criteria set that evaluates the systemic architectural perspective of integrating behavior and user similarity.

\subsection{The CALO program (CALO, PTIME)}

The Cognitive Assistant that Learns and Organizes (CALO) project brought together researchers from 22 organizations to advance research in cognitive software systems by developing a long-lasting, personalized cognitive agent~\cite{sri2021calo}. 
The Defense Advanced Research Projects Agency (DARPA) funded it under the Perceptive Assistant that Learns (PAL) program.
This project led to the CALO meeting assistant as well as a variety of related cognitive agents. 
One example developed to build on the foundation of CALO is PTIME~\cite{berry2011ptime}. 
With its ability to reason, plan schedules, and learn user preferences, PTIME meets the cognitive ability criteria set out in our methodology at a basic level. 
However, due to the implementation of the system with a user interface, it lacks the requested user similarity.
The CALO meeting assistant integrates multiple previously developed cognitive concepts and possesses various cognitive abilities. 
It is the predecessor of Apple Siri.
Nevertheless, in this study context, such systems are classified as conversational user interfaces rather than autonomous user agents.

\subsection{Lebiere et al. 2015 and Wendt et al. 2018}

Cognitive architectures can be directly applied to solve cognitive tasks previously performed by humans.
In Lebiere et al. 2015, the cognitive architecture ACT-R is used to identify malware tasks~\cite{lebiere2015functional}. 
The corresponding model was trained with historical malware data. 
While the agent system can perform its task by employing ACT-R, no other cognitive abilities were apparent. 
Additionally, the system must be initiated manually on the data, not meeting the respective criteria.
Wendt et al. 2018 present a cognitive model capable of building energy management~\cite{wendt2018usage}. 
To achieve that, they employed the KORE cognitive architecture~\cite{zucker2016cognitive}. 
The agent is capable of being integrated externally into a building management interface. 
It processes data from physical sensors and establishes management rules based on that data. 
Like Lebiere et al.'s system, the agent does not employ all of the cognitive abilities in the criteria. 
However, some criteria, such as user similarity, are satisfied with the active perception of the environment and the decision-making.

\subsection{World of Bits (WoB)} 

WoB is an experimental learning platform to train software bots in open-domain web environments~\cite{shi2017world}. 
Agents in WoB perceive the web environments in the form of the Document Object Model (DOM) and rendered pixels.
Interestingly, though benchmark results are low compared to human users, the approach accomplished some web tasks by sending mouse and keyboard actions. 
Furthermore, agents are separated from the environments in which they interact. 
Tasks and activities are low-level operations and lack high-level cognitive capabilities. 
As a result, it falls short of the evaluation criteria.

\section{Discussion}\label{discussion}

The study's results have important implications for the field of cognitive architectures and user-like software bots. 
Of the eleven agents and applied cognitive architectures examined in the previous section, Sec.\ref{results}, only one satisfied the established criteria. 
Interestingly, the agent, IDA by Franklin et al.~\cite{franklin1998ida}, was developed a little over two decades ago. 
Except for the first two candidates, which both precede the year 2000, all the systems featured missed the second criterion of user similarity. The two systems satisfying the criteria were limited by their time and would have been implemented as standalone systems if they had been designed a few years later. 
Presumably, due to a lack of alternatives, they both use email to communicate with their environment. This limitation likely led to them satisfying the user similarity criteria, not necessarily through the designer's intent. 
The trend shown by the other candidates moved towards applications without user similarity. 

\subsection{Bots and Cognitive Architectures}

One central question arising from the results is the absence of software bots currently employing cognitive architectures. 
To the best of our knowledge, it appears highly unlikely to find an agent or bot in current use that meets all the criteria established for this study. 
Despite significant advancements in other areas of artificial intelligence, the fact that only one bot, IDA by Franklin et al.~\cite{franklin1998ida}, satisfied all the criteria suggests that the field of user-like intelligent bots is still nascent and in its early stages. 
These findings and observations align with existing literature, which indicates that integrating cognitive architectures into software bots is a complex and challenging endeavor~\cite{vernon2019architect,gidey2023towards}. 

However, these results also underscore the potential that cognitive architectures hold for enhancing the capabilities of software bots. 
From a simplified architectural perspective, the example of IDA, which met all the criteria, demonstrates how a cognitive architecture can enable advanced autonomous behavior and user similarity in software bots.

These findings also have practical implications. 
These findings highlight the need for developers and researchers to focus their efforts on integrating cognitive architectures into software bots. 
For industry professionals, the results can guide the development of more advanced software bots that align with the requirements of cognitive automation, enabling more effective automation of various tasks across different domains.


\subsection{Autonomy and Behavior: Architectural Perspectives}

The architectural aspects that can address the observed shortcomings in this study, particularly regarding the level of autonomy and generalized behavior, can be viewed from two perspectives: the strict separation of bots and their environment and the integration of a separate behavior model~\cite{gidey2023towards}.

First, as mentioned earlier, the strict separation of agents and their environment facilitates architectural possibilities for autonomy~\cite{gidey2023towards}. 
By decoupling bots from a single environment, they can gain the ability to interact with multiple environments dynamically, achieving higher levels of variability and adaptability. 
This separation also enables the design of software bots with a user-like orientation, treating bots as if they were human users. 
Consequently, this orientation opens up possibilities for alternative architectural design patterns, allowing software bots with user-like characteristics to seamlessly integrate into existing user interfaces or operate on digital platforms without the need for APIs or integration protocols. 
The separation already places the bots in a position where they can potentially establish their own intentions, goals, and deliberate interactions, access, and actions.

Second, an agent's capacity for self-awareness, contextual awareness, and other high-level cognitive functions arises from holistic behavior models~\cite{gidey2023towards}. 
These behavior models are typically formulated using cognitive architectures or similar integrative models and architectures. 
Consequently, the componentization and integration of cognitive architectures into the system architecture of software bots become essential architectural considerations. 
The scientific understanding of machine intelligence, along with related models and principles, is still evolving and has not yet reached a crystallized state~\cite{butlin2023consciousness}.
As a result, models of intelligent behavior will evolve over time, and older ones may require changes and modifications. 
Componentization can facilitate separation and change. 
Additionally, through this componentization, bots can potentially dynamically integrate separate behavior models in real-time.

Adopting these architectural perspectives can contribute to addressing the challenges related to autonomy and generalized behavior in bots, enabling them to operate with greater flexibility, adaptability, and user-like characteristics.

\subsection{Study Limitations}

Despite the study's important findings and insights into integrating cognitive architectures into user-like software bots, the study has limitations. 
The review was limited to publicly available literature and may not capture all existing software bots that employ cognitive architectures. 
Furthermore, the criteria for evaluating software bots may not encompass all possible features and capabilities. 
Future research could expand on our work by investigating databases further, employing different evaluation criteria, or examining the evolution of cognitive architectures in software bots over time.

\section{Conclusion}\label{conclusion}

The review highlights the importance of developing user-like software bots. These bots integrate cognitive architectures, enabling advanced autonomous behavior and user similarity. 
These requirements have significant implications for the engineering aspects of cognitive bots, particularly in cognitive automation.

Through the analysis of architectural recommendations and perspectives, this study has provided insights into achieving these goals at a design level. 
The distinctive view of bots and their environment, along with the dynamic integration of componentized behavior models, arise as key approaches to support both desiderata.

Consequently, implementing these architectural approaches can lead to increased autonomy and adaptability in software bots.
This opens up new possibilities for developing autonomous user-like bots that can effectively perform a wide range of tasks.
These findings contribute to the growing understanding of how to enhance the cognitive capabilities of software bots through architectural references.

\section*{Acknowledgement}
\small{We would like to thank Lorenz Bobber for his valuable support in the initial stages of the survey.}
\addtolength{\textheight}{2.9cm} 
%
%
\bibliography{main}
%
\end{sloppypar}
\end{document}